\documentclass{aa}  

\usepackage{graphicx}
\usepackage{txfonts}
\usepackage{xspace}
\usepackage{natbib}
\usepackage{hyperref}
\usepackage{ mathrsfs }

\newcommand{\mw}{multi-wavelength\xspace}
\newcommand{\g}{$\gamma$\xspace}
\newcommand{\rxj}{RX J1713.7$-$3946\xspace}

\newcommand{\hess}{HESS\xspace}
\newcommand{\fermi}{\textit{Fermi}-LAT\xspace}

\newcommand{\degree}{\ensuremath{^\circ}\xspace} 
\newcommand{\src}{HESS J1731$-$347\xspace}
\newcommand{\vela} {RX J0852.0$-$4622\xspace}
\newcommand{\psr} {PSR\,J0855$-$4644\xspace}

\begin{document}

\title{Study of TeV shell supernova remnants at gamma-ray energies}
\author{
F.~Acero$^{(1)}$ \and 
M.~Lemoine-Goumard$^{(2)}$ \and 
M.~Renaud$^{(3)}$ \and 
J.~Ballet$^{(1)}$ \and 
J.W.~Hewitt$^{(4,5)}$ \and 
R.~Rousseau$^{(2)}$ \and 
T.~Tanaka$^{(6)}$
}
\authorrunning{Acero et al.}

\institute{
\inst{1}~Laboratoire AIM, CEA-IRFU/CNRS/Universit\'e Paris Diderot, Service d'Astrophysique, CEA Saclay, 91191 Gif sur Yvette, France\\ 
\inst{2}~Centre d'\'Etudes Nucl\'eaires de Bordeaux Gradignan, IN2P3/CNRS, Universit\'e Bordeaux, F-33175 Gradignan Cedex, France\\ 
\inst{3}~Laboratoire Univers et Particules de Montpellier, Universit\'e Montpellier 2, CNRS/IN2P3, Montpellier, France\\ 
\inst{4}~Department of Physics and Center for Space Sciences and Technology, University of Maryland Baltimore County, Baltimore, MD 21250, USA\\ 
\inst{5}~Center for Research and Exploration in Space Science and Technology (CRESST) and NASA Goddard Space Flight Center, Greenbelt, MD 20771, USA\\ 
\inst{6}~Department of Physics, Graduate School of Science, Kyoto University, Kyoto, Japan\\ 
\email{fabio.acero@cea.fr} \\
\email{lemoine@cenbg.in2p3.fr} \\
}

\date{}

\abstract 
{ The breakthrough developments of Cherenkov telescopes in the last decade have led to angular resolution of 0.1$\degree$ and an unprecedented sensitivity.
This has allowed the current generation of Cherenkov telescopes (H.E.S.S., MAGIC and VERITAS) to discover a population of supernova remnants (SNRs) 
radiating in very-high-energy (VHE; E $>$ 100 GeV) \g-rays. A number of those VHE SNRs exhibit a  shell-type morphology spatially coincident with the  shock front of the SNR.}
{The members of this VHE shell SNR club are \rxj, \vela, RCW 86, SN 1006, and \src. The latter two objects have been poorly studied in high-energy (HE; 0.1 $<$ E $<$ 100 GeV) \g-rays and need to be investigated in order to draw the global picture of this class of SNRs and to constrain the characteristics of the underlying population of accelerated particles.}
{Using 6 years of \fermi P7 reprocessed data, we studied the GeV counterpart of the SNRs \src and SN 1006. The two SNRs are not detected in the data set and given that there is no hint of detection, we do not expect any detection in the coming years from the  SNRs. However in both cases, we derived upper limits that significantly constrain the \g-ray emission mechanism and can rule out a standard hadronic scenario with a confidence level $> 5 \sigma$.}
{With this \textit{Fermi} analysis, we now have a complete view of the HE to VHE \g-ray emission of TeV shell SNRs. All five sources have a hard HE photon index ($\Gamma < 1.8$) suggesting a common scenario where the bulk of the emission is produced by accelerated electrons radiating from radio to VHE \g-rays through synchrotron and inverse Compton processes.  In addition when correcting for the distance, all SNRs show a surprisingly similar \g-ray luminosity supporting the idea of a common emission mechanism. While the  \g-ray emission is likely to be leptonic dominated at the scale of the whole SNR, this does not rule out efficient hadron acceleration in those objects.}
{}

\maketitle

\section{Introduction}

The breakthrough developments of Cherenkov telescopes in the last decade have enabled
exploration in the very-high-energy (VHE; E $>$ 100 GeV) \g-ray sky with
 angular resolution better than 0.1$\degree$  and an unprecedented sensitivity.
Observations with the current generation of Cherenkov telescopes (H.E.S.S., MAGIC and VERITAS) of the Galactic plane have 
revealed a population of supernova remnants (SNRs) radiating in VHE \g-rays.
Such radiation is the signature that particles are accelerated to multi-TeV energies in SNRs.

VHE observations, combined with longer wavelength information, have been used to probe the nature (hadrons or leptons)
and the energy properties (energetics and maximum energy) of those accelerated particles in order to 
investigate  the origin of Galactic cosmic rays (CRs).

The list of VHE sources associated with a SNR has been steadily growing in 
the last decade and a dozen detections have been reported so far (see, for example, the TeV online catalog TeVCat\footnote{http://tevcat.uchicago.edu} or 
the catalog of high-energy observations of Galactic SNRs\footnote{http://www.physics.umanitoba.ca/snr/SNRcat/, \citet{ferrand12} }).
This sample  could be described in two main categories: 1) a population of young objects ($ t \lesssim$ 5 kyrs)  where the emission is likely associated 
with the SNR shell and 2) a group of  older SNRs ($ t > $ 10 kyrs)  that are mostly radiating through their interaction with the surrounding molecular 
clouds (e.g. W28, IC443).

For a number of objects in the first group with large angular size ($R_{\rm SNR}> 0.25\degree$), 
the VHE emission is spatially resolved and exhibits a shell-type morphology spatially coincident with the shock front  of the SNR.
The members of this VHE shell SNR club are \rxj \citep{ah04}, \vela \citep{ah07-VelaJr},  RCW 86 \citep{ah09},   SN 1006 \citep{acero10}, and \src \citep{acero11b}.
In those objects the \g-ray emission allows us to probe the population of high-energy particles directly at the shock where the acceleration is taking place.
Those objects also share a number of similarities. For example, they are evolving in low-density ambient medium  ($<$ 1 cm$^{-3}$) and are bright X-ray synchrotron emitters (see Sect. \ref{similarities} and Table \ref{params} for references).

The VHE emission can be produced either from the Inverse Compton (IC) scattering of electrons off the ambient photon field or in interactions of hadrons with ambient matter.
Understanding the nature of the \g-ray emission based solely on the VHE observations is a difficult task as both mechanisms produce
similar spectra in the VHE regime.
The Large Area Telescope (LAT) onboard the \textit{Fermi Gamma-ray Space Telescope} is operating in the high-energy (HE; 0.1 $<$ E $<$ 100 GeV) \g-ray domain,
 a crucial energy range where the spectral signatures of the leptonic and hadronic scenario differ. The hadronic emission
in a case with an $E_{\rm p}^{-2}$ proton spectrum produces approximately an $E_{\gamma}^{-2}$ spectrum in the GeV regime\footnote{Assuming that the cutoff energy in the particle population is beyond the HE regime.}. 
 In the case of IC emission, an $E_{\rm e}^{-2}$ electron population  translates into a spectrum with a slope $E_{\gamma}^{-1.5}$.

Joint studies of the HE/VHE emission in a \mw context (including radio and X-ray observations) are providing new means to disentangle  the different scenarios.
The best example so far is the case of \rxj, one of the brightest SNRs in the VHE \g-ray sky and considered as one of the prototypes of
an efficient CR accelerator. The observations with \fermi of \rxj have revealed a hard spectrum at HE with a photon index $\Gamma$=$1.50\pm0.11_{\rm stat}$
\citep{abdo11}, which tends to be incompatible with purely hadronic models \citep[e.g.][]{ellison10,zirakashvili10}. A similarly hard index 
 \citep[$\Gamma \sim 1.4 \pm 0.2_{\rm stat}$; ][]{yuan14} was found for RCW 86. 

In the case of \vela, HE \g-rays were detected in spatial coincidence with the SNR and the resulting HE spectrum is found 
to be hard (although systematic errors are rather large) with a spectral index of   $\Gamma = 1.85 \pm 0.06_{\rm stat} \pm 0.18_{sys} $ \citep{tanaka11}.
However, we note that a fraction of the HE emission seen by \fermi from \vela could be associated with the pulsar wind nebula (PWN)
 surrounding \psr, which would harden the spectral slope.
This PWN is  in spatial coincidence with the south-eastern shell of the SNR and is seen in X-rays 
\citep{acero13} and possibly VHE \g-rays \citep{pazarribas12}.

For \src and SN 1006, no \fermi source is listed at the position of the two SNRs in the third \textit{Fermi} catalog \citep[3FGL\footnote{The 3FGL uses 4 years of P7 reprocessed data.}, ][]{3FGL}.
SN 1006 has been studied using 3.5 years of \fermi P7V6  data by \citet{araya12} and no detection was reported. An
upper limit was reported (assuming a point source) on a large energy range (500 MeV to 100 GeV) that is not well optimized to constrain the models.
The last object of the VHE SNR club, \src, was studied in the HE domain with 3.5 years of \fermi P7V6 data by \cite{yang14} and no detection was reported.

With the advent of the new 
 \fermi reprocessed data (P7REP), which provide an improved
sensitivity and angular resolution (see Sect. \ref{fermi}) together with a new Galactic diffuse model,
 we  investigate with 6 years of data the HE counterpart of the two TeV shell SNRs (SN 1006 and \src) that have not yet been detected.
 In Sect. \ref{fermi} we present the data analysis while in Sect. \ref{discussion} we discuss the implication of the results with respect to the nature of the \g-ray emission
 and investigate the general class properties of the TeV shell SNRs.  \\

\section{\g-ray observations with \fermi}
\label{fermi}

\subsection{Data analysis}

The \textit{Fermi} Large Area Telescope (LAT) is a $\gamma$-ray telescope operating from 20 MeV to energies greater than 300 GeV.
A description of the instrument and of its performance is presented in \citet{atwood09}.
In operation since August 2008, the LAT  provides the most sensitive all-sky survey in the HE \g-ray regime \citep{ackermann12LAT}.

The analysis presented here was carried out with  6 years of data from August 4, 2008  to August 4, 2014 using the LAT reprocessed Pass 7 data (P7REP).
A detailed description of the reprocessed data can be found in  \citet{bregeon13}.
This reprocessing results in several improvements in the quality of LAT data. Among them,
 the point-spread function (PSF) is significantly improved above a few GeV ($\sim25$\% smaller above 10 GeV) and the significance of detection and precision of measured photon flux is increased slightly for most sources, more strongly for sources with hard spectra as  could be the case for some young SNRs. 

The P7REP  \texttt{SOURCE} class events and the Instrument Response Functions (IRFs)  \texttt{P7REP\_SOURCE\_V15}  were used for this study.
We selected only events with energies greater than 3 GeV as a compromise between statistics and background from the diffuse Galactic emission, and
zenith angles smaller than 100$\degree$ to reduce contamination from  the Earth limb \citep{abdo09-limb}.
Time intervals when the rocking angle was more than 52$\degree$ and when the \textit{Fermi} satellite was within the South Atlantic Anomaly were also excluded.
The Fermi Science Tools\footnote{http://fermi.gsfc.nasa.gov/ssc/data/analysis/}  v9r31p1 and the corresponding Galactic diffuse background (\textit{gll\_iem\_v05\_rev1.fit}) and  extragalactic isotropic background (\textit{iso\_source\_v05.txt}), were used.

In addition to the diffuse backgrounds, the model of the region of interest contains the 3FGL sources within a 10$\degree$ radius of the SNRs positions.
Using this model, we computed residual test statistic (TS) maps to search for additional background sources. 
The residual TS map is obtained by computing the TS value for an additional point source at each point of the grid in excess of a given model for the region of interest
 \citep[see Sect. 3.1 of ][ for a more detailed description of the concept of residual TS maps]{nolan12}.
The positions of the excesses (with a threshold in residual TS of 25) were used as seeds to define the position of additional \g-ray sources not present in 3FGL. 

With 6 years of P7REP data, we investigated  the GeV counterparts of those two SNRs using a binned likelihood fit 
on a $7\degree \times 7\degree$ region of interest. Both SNRs are approximately 0.5\degree in diameter and, for the energy range considered here,  should
 be considered as spatially extended sources for \fermi .
Therefore they  were modeled  using morphological templates derived from the \hess excess maps. 
The energy threshold for these excess maps is 240 GeV  and 500 GeV for \src and SN 1006 respectively \citep{acero10,acero11b}.

Although SN 1006 is located at high Galactic latitude ($\ell = 14.6\degree$) and is not as affected by the Galactic diffuse emission as \src,
the same energy threshold ($E > 3$ GeV) was used for both SNRs for the sake of comparison.

\subsection{Results: \src}
\label{results1}

Using the model presented in the previous subsection (diffuse models+3FGL sources), we generated a TS map of the region to search for new background sources.
Based on the TS map, two additional background point sources (PS-SE and PS-NW, see  Fig.\ref{TSmaps}) at the Galactic positions $\ell, b$ = (354.34\degree, -1.11\degree), 
(352.93\degree, -0.30\degree) were detected with TS=41.19, 31.22 respectively.

 We generated the residual TS map with a model that includes these 2 new sources but not SNR \src. This residual TS map is shown in Fig.\ref{TSmaps} along with the positions of  the 3FGL sources and the newly added background sources. 
No significant excess \g-ray emission is observed in the region of the SNR. Within the SNR contours, 
the maximal TS value is 6.4. 
s
To investigate the \g-ray emission under the extended source assumption, we added the \hess template to the model using a power-law spectral model.
The TS value of the SNR derived from the binned likelihood fit  is 2.58/2.55/2.28 assuming a spectral index of 1.5, 2.0 and 2.5 respectively.

To derive upper limits for this extended source, we used the \hess morphological template of the SNR only (the nearby source HESS J1729$-$345, 
contours shown in Fig. \ref{TSmaps}, is not included). 
A power-law spectral distribution with a fixed spectral index of 2.0 was assumed. Upper limits at 95\% confidence level (CL)  were derived using a Bayesian method implemented in the Python tool \textit{IntegralUpperLimit} provided in the Fermi ScienceTools.
The energy bands 3--30 GeV and 30--300 GeV were chosen as the best compromise between statistics and contamination from the diffuse Galactic emission.
The resulting limits (listed in Table \ref{TStable}) are shown  in Fig. \ref{SED} together with the \hess spectral data points.
In the most constraining energy bin (3--30 GeV), changing  the  spectral index to 1.5 or 2.5 alters the results by 10 to 20\% (see Table \ref{TStable}).

While the analysis was carried out for E > 3 GeV  in order to minimize the impact of the Galactic diffuse emission,
its contribution in the Galactic plane is non-negligible even at these high energies.
 To estimate the associated systematic uncertainty we applied the procedure used in several \textit{Fermi} analyses of Galactic sources
\citep[e.g. the second \fermi catalog of \g-ray pulsars, ][]{abdo13-2PC}. In this procedure the best-fit value of the normalization of the Galactic diffuse component
 is increased/decreased by 6\%\footnote{  The specific value of 6\% represents
the 1.5$\sigma$ deviation in the distribution of Galactic diffuse normalization parameters from all the fits
 in the second catalog of pulsar \citep{abdo13-2PC}.}, and the flux is recomputed with the latter normalization parameter kept frozen.
 As we are dealing with upper limits, we  decreased the normalization of the Galactic diffuse
 emission by 6\% (the residuals are thus higher), froze this parameter,
 and recomputed the upper limit. The resulting upper limit is 1.68 $\times 10^{-12}$ erg cm$^{-2}$ s$^{-1}$
 in the 3--30 GeV energy range   (a $\sim$2\% increase in comparison with the value in Table \ref{TStable}) 
  while the upper limit in the 30--300 GeV energy bin was unchanged.
 
\subsection{Results: SN 1006}

\begin{figure}[t]

\includegraphics[bb= 40 120 575 630,clip,width=8.5cm]{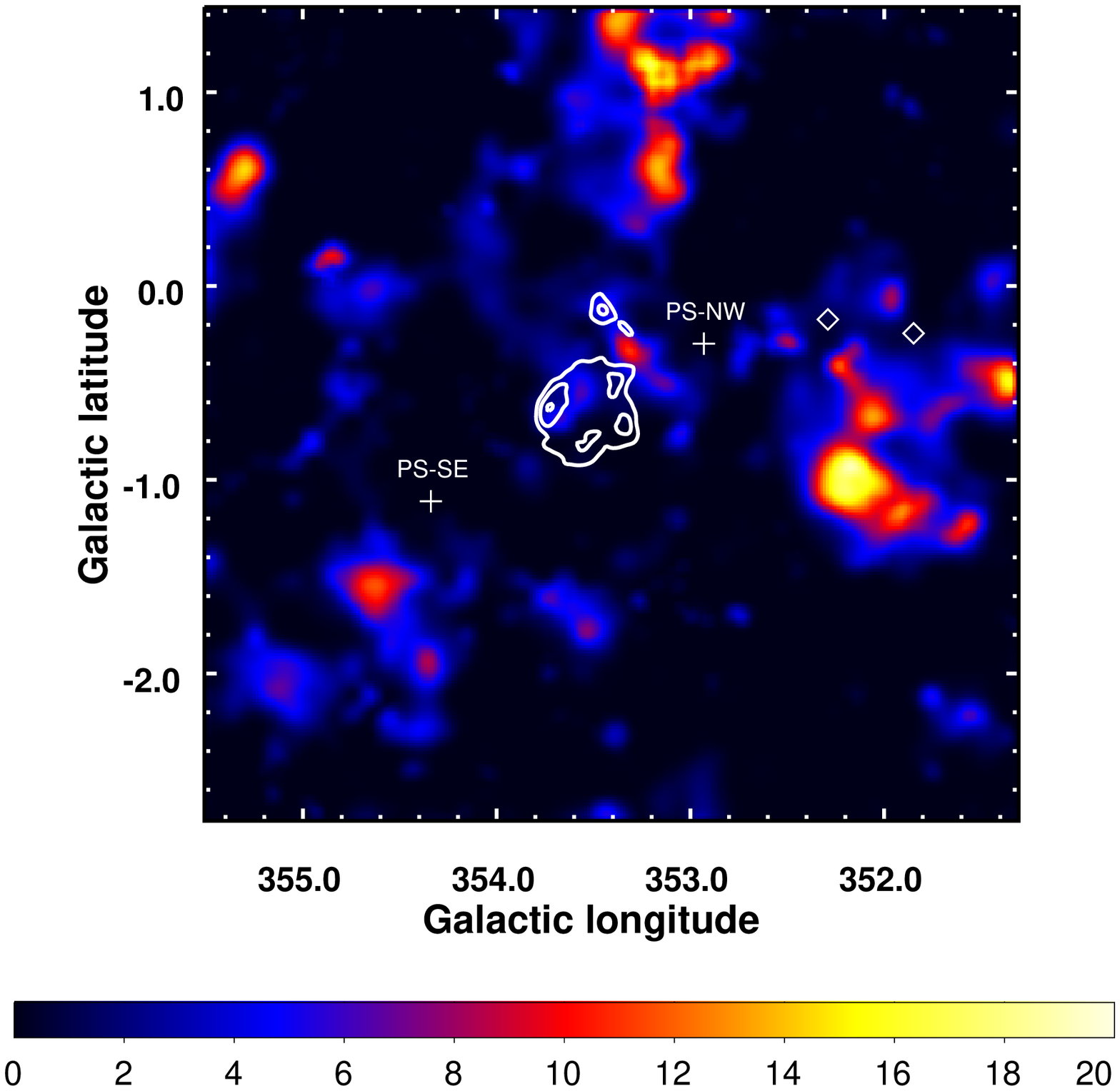}
\includegraphics[bb= 40 120 575 630,clip,width=8.5cm]{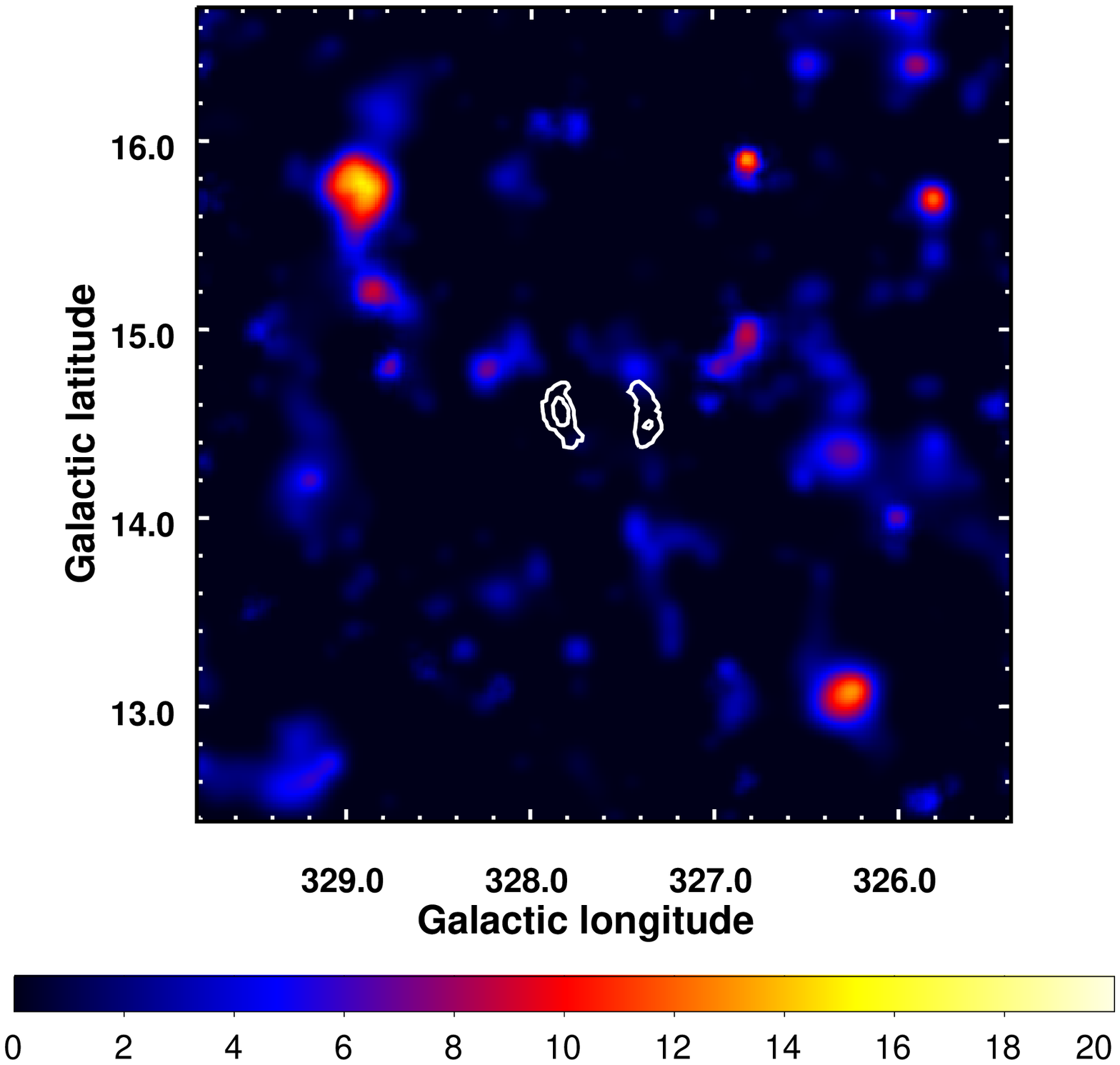}

\vspace{-0.3cm}
\caption{
\footnotesize
Residual test statistic (TS) maps for E $>$ 3 GeV in a $4^\circ \times 4^\circ$ region for \src (top) and SN 1006 (bottom). The background model is described in Sect. \ref{fermi} and the SNRs are not included.  The point sources included in the model are represented by diamonds (3FGL sources) and crosses (additional background sources). The source HESS J1729$-$345 seen near the SNR \src is shown on the TS map (top) but is not included the morphological template used to derive upper limits. No \fermi sources are detected in the vicinity of SN 1006. }
\label{TSmaps}
\end{figure}

The same procedure, described in section \ref{results1}, is applied to the region of SN 1006. 
No  background sources were detected on top of the standard model (diffuse emission+3FGL sources). 
The resulting residual TS map in shown in Fig. \ref{TSmaps} and no significant emission is detected within the contours of the SNR. 
Using the \hess morphological template of SN 1006, the TS value from the likelihood fit is very small (TS=0.95). 
Due to the bipolar morphology of SN 1006, both limbs were tested independently by splitting the \hess template as two regions.
As both regions are separated by $\sim$0.5\degree, the \fermi PSF above a few GeV is sufficient to disentangle the contribution from both limbs.
The resulting TS from  the north-eastern and south-western limbs are respectively 0.13 and 0.91 for a spectral index of 2.0. 

\begin{figure}[t]

\includegraphics[bb= 15 15 455 380,clip,width=8.5cm]{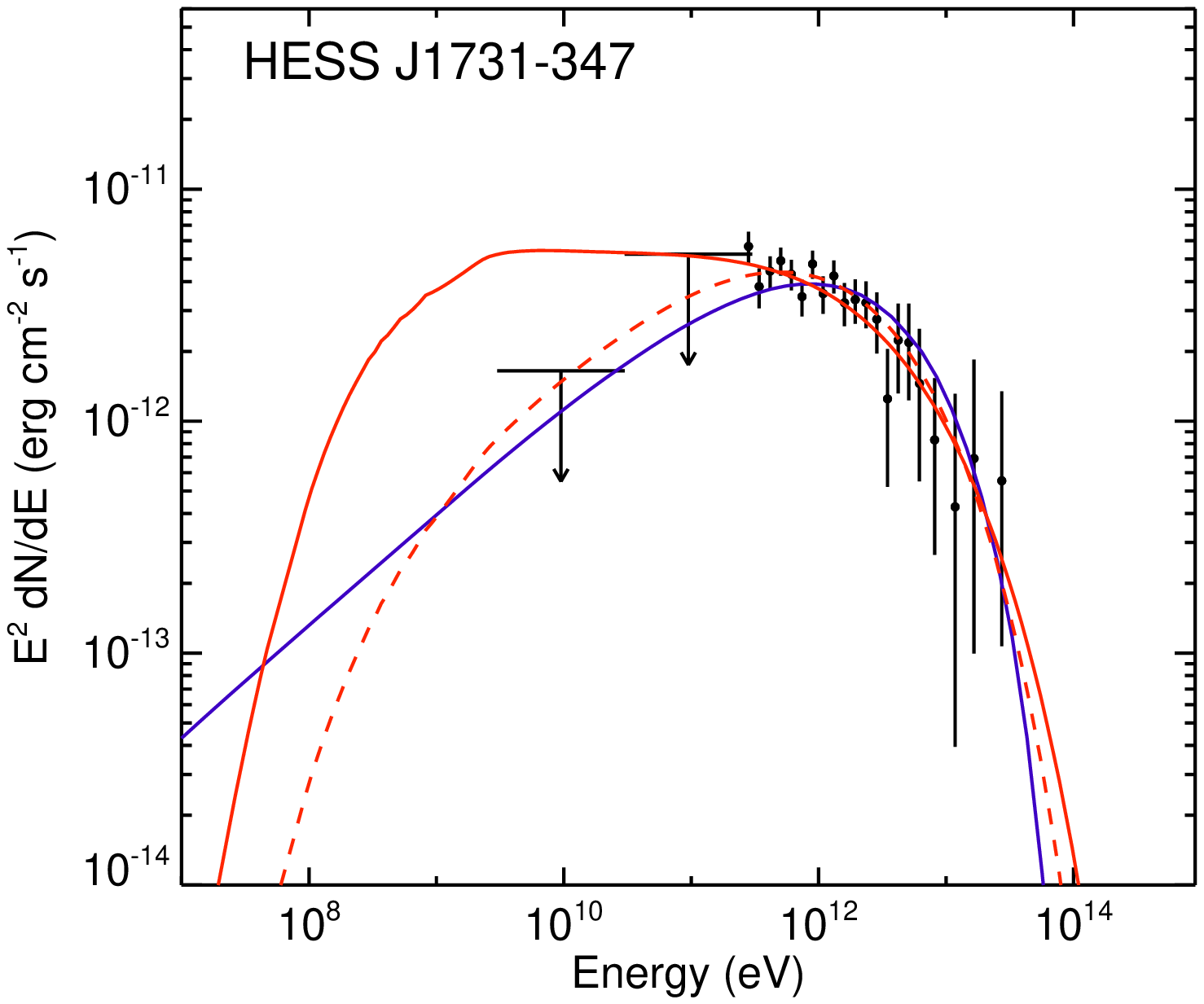}
\includegraphics[bb= 15 15 455 380,clip,width=8.5cm]{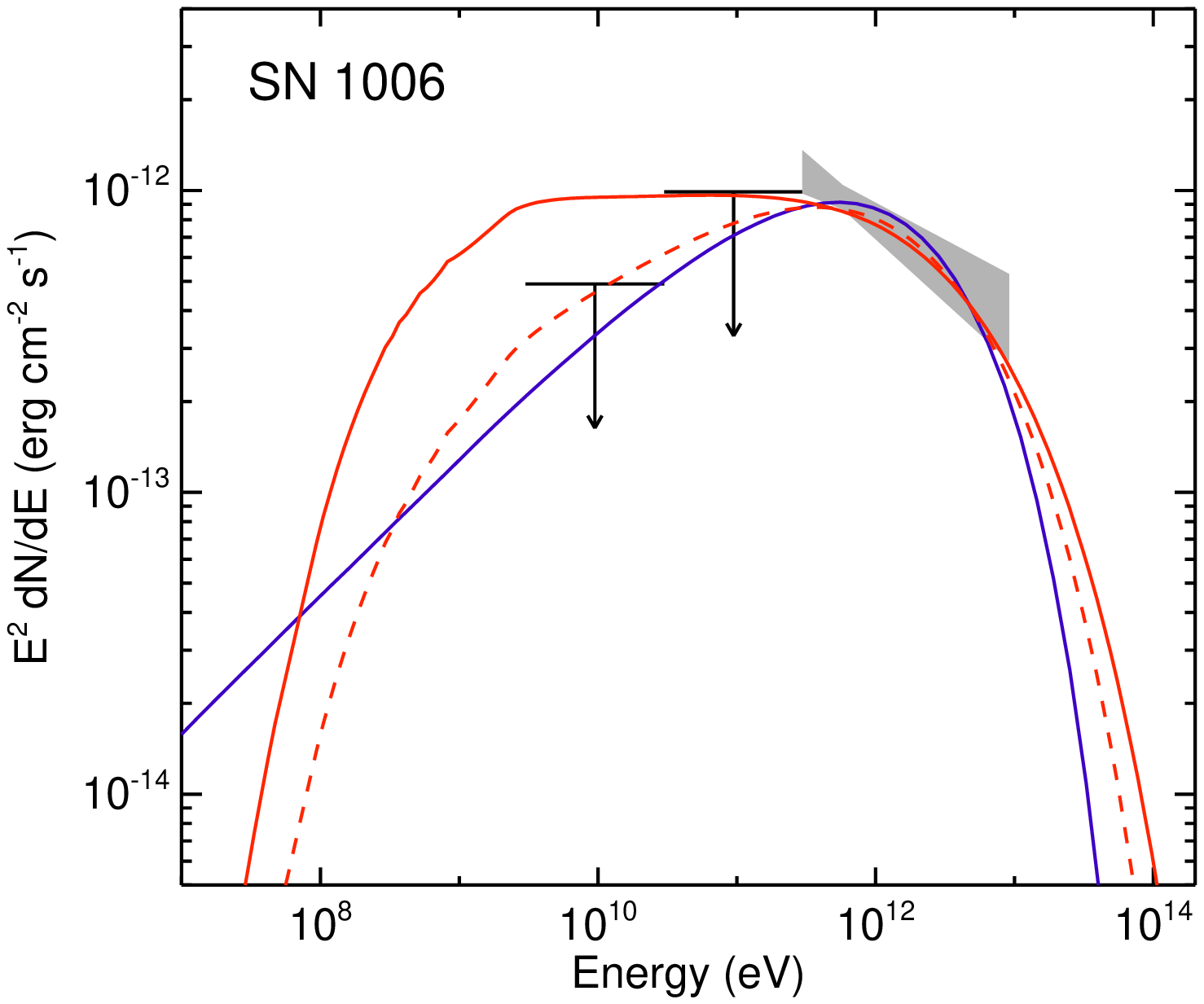}

\vspace{-0.3cm}
\caption{
\footnotesize
Broadband Spectral Energy Distributions (SEDs) for the SNR \src (top) and SN 1006 (bottom).
 A purely leptonic (blue solid) and a hadronic (red solid) scenario are shown in the test particle case ($s_{\rm e,p}$=2.0). The red dashed line shows
  the steepest hadronic  spectral slope compatible with the data, respectively $s_{\rm p}$=1.5 and 1.7 for \src and SN 1006. The HESS data points for the top and bottom panels are extracted
 from \citet{acero11b} and \citet{acero10}  respectively and the \fermi 95\% confidence level  upper limits are derived assuming a spectral index of 2.0.
  }
\label{SED}
\end{figure}

Because of the low significance of SN 1006 using 6 years of P7REP data, we do not expect any significant detection in the coming years. 
Upper limits using the morphological template for the whole SNR and for the two limbs are derived in the 3--30 GeV and 30--300 GeV energy bands and are reported in Table  \ref{TStable}. The limits for the whole SNR together with the \hess data are shown in Fig. \ref{SED}.
As SN 1006 lies 14\degree above the Galactic plane, we expect the contribution from the Galactic diffuse emission  above 3 GeV to be negligible. We applied the same test as in the previous section for a systematic effect  related to the Galactic diffuse emission normalization and the upper limits remained unchanged.

\begin{table}
\begin{center}
\begin{tabular}{l|c|c|c|c}

     Index                            & J1731$-$347     & SN1006    &    SN1006 NE    &    SN1006 SW  \\
               \hline
               \multicolumn{5}{c}{ TS values} \\ \vspace{0.2cm}
2.0   &      2.55  &         0.95  &      0.13    &      0.91   \\

               \multicolumn{5}{c}{ Upper limits } \\
1.5   &    1.36/6.25     &     0.49/1.39      &     0.19/0.55    &     0.19/0.51    \\
2.0   &    1.65/5.21     &     0.47/0.97      &     0.16/0.44    &     0.19/0.35    \\
2.5   &    2.05/4.32     &     0.43/0.71      &     0.13/0.41    &     0.18/0.26    \\

\end{tabular}
\end{center}
\caption{\fermi test statistics (TS) values obtained using the \hess morphological templates. For the case of SN 1006, 
the \hess template was divided into the two bright limbs to test them independently. In the second part of the table, the 95\% confidence level upper limits (in units of 
$10^{-12}$ erg cm$^{-2}$ s$^{-1}$)  are reported for different spectral index in the two energy bands 3--30 GeV and 30--300 GeV.  }
\label{TStable}
\end{table}%

\section{Discussion}
\label{discussion}

To investigate the nature of the  \g-ray emission from both SNRs, we focused on the HE upper limits derived in the previous section and the VHE data
 collected in the literature. The data are then compared with a simple one-zone static model where the particles (electrons and protons) are described by
 a power-law with an exponential cutoff\footnote{We note that in all the cases, except \rxj, the statistics at HE and VHE are not sufficient to measure the 
 shape of the cutoff.} of the form:
 
 \begin{equation}
  \mathrm{d} N_{\rm e,p}/\mathrm{d}E  \propto E^{-s_{\rm e,p}} \times exp(- (E/E_{\rm cut}^{\rm  e,p})^{\beta} ).
\label{expcut}
\end{equation}
The sharpness of the cutoff is represented by the factor $\beta$. In most of the literature, an exponential cutoff is assumed ($\beta$=1.0) which we will use
throughout this study for the sake of comparison. However, we note that a broader cutoff with $\beta=0.5$ can also fairly well reproduce the \mw data 
and in particular the low-energy part of the VHE spectra \citep[see ][ for an example on \rxj]{liu08, li11}.
 
 The \g-ray emission from the $\pi^{0}$ decay is calculated following the method of \citet{huang07} where accelerated protons and helium nuclei collide with the  
 interstellar medium with standard composition. The IC emission is derived from the scattering on the Cosmic Microwave Background (CMB) and the local interstellar infrared radiation
 field  derived from \citet{porter06}. 
A lower limit to the distance of \src of 3.2 kpc  was derived in \citet{acero11b}. In the following discussion the distance is fixed to 3.2 kpc and therefore 
the electrons and protons energy budget ($W_{\rm e}$ and $W_{\rm p}$) should  be viewed as lower limits. The distances to the other SNRs are listed in Table \ref{params}.

\subsection{Constraints on emission model for HESS J1731$-$347}

The \fermi upper limits reported in Section \ref{fermi}  are shown in Fig. \ref{SED} with respect to the \hess data points. 
In a purely hadronic scenario both the HE and VHE emission result from the $\pi^{o}$ decay.
In the test particle case with a proton spectrum index $s_{\rm p} = 2.0 $ (illustrated by a red solid line in Fig. \ref{SED}),  
the purely hadronic scenario is excluded by the \fermi upper limits. In order to quantify to which extent we can exclude this hadronic scenario, we
compared the log-likelihood  ($\mathscr{L}_{\rm 0}$) obtained when the SNR flux is 
frozen to the hadronic case,  to the log-likelihood ($\mathscr{L}_{\rm 1}$) given by the fit of the SNR when the flux is let free. In both cases the log-likelihoods are computed using \textit{Fermi} data, i.e. in the 3--300 GeV range, and the photon spectral index was fixed to 2. For $\mathscr{L}_{\rm 0}$, the SNR flux is frozen to a value of  $10.21 \times 10^{-10}$ cm$^{-2}$ s$^{-1}$ in the 3--300 GeV band which corresponds to the flux in the purely hadronic case where $s_{\rm p}$=2.0 that connects to the \hess data points. 

The resulting $\Delta \mathscr{L}=( \mathscr{L}_{\rm 1} - \mathscr{L}_{\rm 0})$  is 14.68 for one additional degree of freedom and we therefore conclude that the hadronic test particle scenario is excluded at the $\sim$5.4 $\sigma$ confidence level.
In order to accommodate the HE upper limit and the VHE data in a hadronic model, a proton slope $s_{\rm p} \leq $ 1.5 is required  (see dashed line in Fig. \ref{SED}).

In several aspects (age, bright X-ray synchrotron emission, low-density ambient medium, VHE luminosity) \src is very similar to \rxj (see Table \ref{params}). 
Our \fermi analysis has shown that \src has very similar properties at HE compared to \rxj whose measured photon index is 1.5$\pm$0.1. 
It is therefore likely that the dominant emission mechanism in \g-ray for \src is similar to the case of \rxj (i.e. leptonic dominated).

\subsection{Constraints on emission model for SN 1006}

The upper limits derived for SN 1006 by \citet{araya12} were not sufficiently constraining to 
be able to disentangle between the different emission scenarios possible. We also note that SN 1006 was modeled  as a point source in their study.
 While this is a reasonable assumption for \fermi in the 100 MeV energy range, the radius of the SNR (R=0.25\degree) becomes non-negligible in comparison with the PSF 
above a few GeV. In consequence, the value of the upper limits for SN 1006 can be artificially low when the point source hypothesis is used.

Here, with 6 years of P7 reprocessed data and assuming an extended source hypothesis (the HESS template), we show in Fig. \ref{SED} the first constraining upper limits at HE.
As for \src, we compared the log-likelihoods in a scenario where the 3--300 GeV flux is fixed to the  hadronic hypothesis ($2.19 \times 10^{-10}$ cm$^{-2}$ s$^{-1}$,  red line in Fig. \ref{SED}) and a scenario in which the flux normalization is let free. With a $\Delta \mathscr{L}= 17.13$, the hadronic scenario is excluded at a 5.8 $\sigma$ confidence level.
The steepest hadronic spectral slope compatible with the HE upper limit and the VHE data points gives  $s_{\rm p} \leq $ 1.7 (red dashed line in Fig. \ref{SED}).

When using the HESS template to model the southwest rim only of the SNR, we find no evidence of HE emission.
While the synchrotron cut-off frequency is slightly higher in the northeast rim \citep{miceli09}, the southwest  rim is of high interest to study the hadronic scenario
 as this region is the only one in the SNR where  we have evidence of efficient particle acceleration at the shock (traced by the non-thermal X-rays), dense target densities 
 (traced by the HI observations) and proof of the interaction between the shock and the cloud (shock front is curved inwards in this region) as reported by \citet{miceli2014}.
 However this interaction region represents a small angular fraction of the whole SNR and requires detailed HE and VHE spatially resolved 
  spectroscopy that is at the limit of the capacities of current generation instruments.

\begin{table*}
\begin{center}
\begin{tabular}{p{2.25cm}|p{2.25cm}|p{2.25cm}|p{2.25cm}|p{2cm}|p{2cm}}

 & RX J1713  & RX J0852 & HESS J1731 & RCW 86 & SN 1006 \\
               \hline
SN nature  &  CC       &            CC             &      CC                  &             Type Ia        &          Type Ia               \\               
Distance (kpc) &  0.9-1.3        &      0.6-0.9        &           $\geq$ 3.2            &        2.3-2.8           &          2.0-2.4          \\
Radius  \, \,  (pc)  &     10    &    12         &          14            &          15            &           10        \\
Age     \, \, \, \,   (kyrs) &     1.6    &       2-4      &              2-6        &           1.8           &          1         \\
Density \,   (cm$^{-3}$) &  $<$0.02        &      $<$0.03        &           $<$0.01            &        0.1-0.5            &          $<$0.05       \\


$\Gamma_{\rm HE}$ &         1.5 $\pm$ 0.1       &             1.85$\pm$ 0.06            &                $\leq$ 1.5*        &                1.4  $\pm$ 0.2        &             $\leq$1.7*              \\
 $\Gamma_{VHE}$ &  2.32 $\pm$ 0.01  &  2.22 $\pm$ 0.06  &  2.32 $\pm$ 0.06  &  2.41 $\pm$ 0.16  &  2.30 $\pm$ 0.15  \\

References  &  12,  17,  11,  6,  1  &  13,  7,  18,  20, 9  &  21,  3,  22,  5  &  19,  10,  15,  8,  24  &  23,  2,  14,  4,  16  \\

$s_{\rm e}$                   &  2.15          &     2.15                      &        2.02                    &             2.30             &               2.10            \\
E$_{\rm cut}$ \,  (TeV)                 &   51         &         25                  &                   24         &            22              &             10              \\
W$_{\rm e}$ \, (10$^{48}$ ergs) &     0.55       &             0.38              &               0.18             &          1.2                &           0.21                \\

References &  \citet{yuan11}    &     \citet{tanaka11}         &     \citet{yang14}          &     \citet{yuan14}            &        \citet{acero10}         \\

\end{tabular}
\end{center}
\caption{Summary of the physical properties of the known TeV shell morphology SNRs.  The corresponding nature of the SN explosion is noted as core collapse (CC) or thermonuclear (Type Ia).  The ambient medium densities shown in this table have been derived from the presence/lack of X-ray thermal emission. The best-fit spectral indices at HE and VHE are reported under a power-law hypothesis in the corresponding energy band and the errors reported are statistical. In the second part of the table, the best-fit leptonic model for each SNR is reported. The electron energy budget (W$_{\rm e}$) is given for E$_{\rm e} >$ 1 GeV. For \vela and SN 1006,  W$_{\rm e}$  was given for E $>$ 10 MeV (respectively 100 MeV) and the values in the table have been rescaled.
 References: 1) \citet{abdo11},  2) \citet{acero07},  3) \citet{acero09b},  4) \citet{acero10},  5) \citet{acero11b},
   6) \citet{ah07},  7) \citet{ah07-VelaJr},  8) \citet{ah09},  9) \citet{allen15},  10) \citet{Bocchino:2000rm},  11) \citet{cassam04},
     12) \citet{Fesen:2012fk},  13) \citet{katsuda08},  14) \citet{katsuda09},  15) \citet{lemoine12},
       16) \citet{miceli12},  17) \citet{Moriguchi:2005fj},  18) \citet{pazarribas12},  19) \citet{sollerman03},
         20) \citet{tanaka11},  21) \citet{tl08},  22) \citet{tl10},  23) \citet{winkler03},  24) \citet{yuan14}. \newline 
 *: this work. }
\label{params}
\end{table*}%

\subsection{Source Class comparison at HE and VHE}
\label{similarities}

In order to understand the underlying emission mechanism at HE and VHE, we investigate and compare the \g-ray properties of the members of the class.
Among those properties, the photon spectral index observed at HE is a key ingredient to disentangle between a test particle leptonic/hadronic scenario.
The HE and VHE spectral index of the five SNRs considered here are listed in  Table \ref{params}.
We note that in a simple one zone leptonic model, the slope of the electron population is in part constrained from the radio to X-ray synchrotron radiation. 
However, as there are usually no data between the radio and the X-ray data points, the slope and the synchrotron cutoff frequency 
can be degenerate parameters in particular if synchrotron cooling effects are significant and if curvature is allowed in the particle spectrum.
In the case of \rxj,  \vela and RCW 86 which were detected in the \fermi data, the electron slope can actually be fitted over a larger energy range using  GeV to TeV data,
thus giving better leverage on the electron slope and reducing the impact of the aforementioned caveats.
In the two cases where \fermi upper limits are reported, the electron slope can be regarded as an upper limit (i.e. the steepest slope allowed by the \fermi upper limits).

For the five SNRs considered, all show, or are compatible with, hard HE spectral indices ($ 1.4 < \Gamma < 1.8$) that exclude the standard hadronic
test particle scenario. All photon indices (except for \vela) are compatible with a test particle leptonic dominated scenario where the electron slope
is $s_{\rm e} = 2.0$, which translates into photon spectral index of 1.5. In the case of \vela, the slightly higher HE photon index
  \citep[1.85$\pm 0.06_{\rm stat} \pm 0.18_{\rm syst}$, ][]{tanaka11}   could be due to a deviation from the test particle case, a mix of hadronic and leptonic contributions 
  or a possible contamination  from the pulsar wind nebula seen around \psr \citep{acero13} that is located right on the south-eastern part of the SNR shell.

Theoretical possibilities to explain a hard spectral index ($\Gamma_{\rm HE} < 2.0$) within a hadronic scenario  include  back reaction 
effects \citep{bv06,zirakashvili10} or shock-cloud interaction \citep{inoue12,gabici14}. However in young
SNRs where the hadronic hypothesis is preferred, the  measured spectral indices are softer or equal to 2.0 as in Cassiopeia A \citep[2.0$\pm 0.1_{\rm stat} \pm 0.1_{\rm syst}$, ][]{abdo10-casA}  and Tycho \citep[2.3$\pm 0.2_{\rm stat} \pm 0.1_{\rm syst}$, ][]{giordano12}.

This similarity of hard photon spectral indices in our SNR  sample tends to point towards a common leptonic dominated scenario for the HE and VHE \g-ray emission.
We note that while this is probably true when looking at the spectrum averaged on the whole SNR, there could be some smaller subregions (e.g. dense clumps) 
where the hadronic mechanism could significantly contribute to the local \g-ray emission.

\begin{figure}

\includegraphics[bb= 15 15 455 380,clip,width=8.5cm]{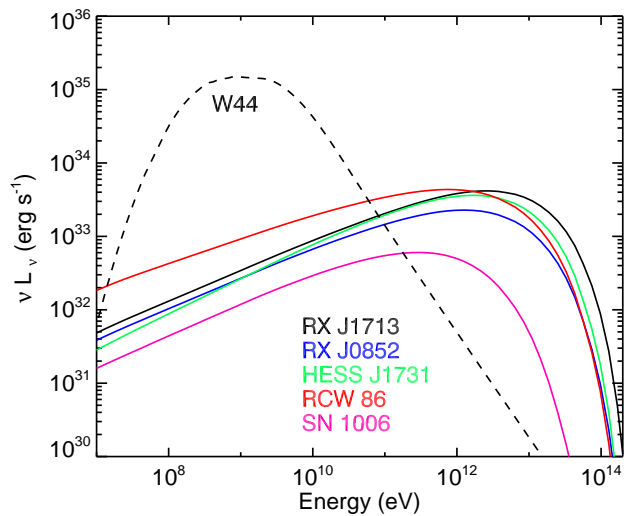}

\vspace{-0.3cm}
\caption{
\footnotesize
Intrinsic spectral energy distribution for all the members of the TeV shell SNR club in the GeV-TeV energy range. 
For the sake of comparison the SED from the SNR W44 (age$\sim$ 20 kyrs),  where the evidence for hadronic emission is secure,  is shown by the dashed line. The corresponding parameters used to produce this figure are shown in Table \ref{params} except for W44 where the broken power-law model from \citet{ackermann13-pi0} is used.}
\label{SEDlum}
\end{figure}

Based on the model parameters compiled  from the leptonic model fits in the literature, the SEDs in HE and VHE \g-rays
in luminosity space of the five shell SNRs are presented in Fig. \ref{SEDlum}.
In addition to a similar HE spectral index, 
this comparison plot reveals a striking similarity in terms of peak luminosity and spectral shape for the SNRs considered in this sample.
This similarity is highlighted when compared with the SED of the SNR W44 where the evidence for hadronic emission is secure 
\citep[detection of the $\pi^{o}$ decay feature, the smoking-gun evidence for hadronic emission, in the $<$ 100 MeV energy range: ][]{ackermann13-pi0}.
The \g-ray luminosity of SN 1006 in Fig. \ref{SEDlum} is lower than for other SNRs. This is probably  related to the SNR bipolar morphology and the reduced surface for efficient  particle acceleration. If we correct for this effect by a renormalization factor of 0.2 as discussed in \cite{berezhko09}, the peak luminosity is comparable to other SNRs.

This similar \g-ray luminosity in Fig. \ref{SEDlum} was not a priori expected given the fact that our sample is composed of different types of 
SN explosion and ages ranging from 1 to 6 kyrs (see Table \ref{params}).
Nevertheless,  the sources in our sample share an important  characteristic that could explain part of this \g-ray similarity. This characteristic is that the
 SNRs have evolved for most of  their life in a low-density ambient medium  (see Table \ref{params}) which allowed them to maintain a high shock velocity
  over a long period of time and therefore efficiently accelerate particles to high energies.
   In this low-density ambient medium, the \g-ray emission is dominated by the leptonic mechanism in which the main source of the photon field, the CMB, is common to all the SNRs. 
    We note that this low-density argument is likely an over-simplification in the case of RCW 86.  This object probably exploded in a low-density  cavity 
    blown by the wind of a single degenerate system, and only recently have parts of the shock  started  to interact with the border of the cavity \citep{williams11,broersen14}. 
    As a result, the thermal X-rays (tracing the shocked ISM) and the non-thermal X-ray emission (tracing the high-energy particles) do not stem exactly from the same region. 
     A similar caveat applies in the case of \rxj and \src, which are surrounded by molecular clouds \citep{fukui12,fukuda14} with possible interactions in some regions of \rxj. 
     
It is surprising that RCW 86, where part of the shock emits thermal X-rays tracing ambient medium densities of $\sim$0.5 cm$^{-3}$,
 shows a similar HE spectral index to \rxj where no thermal X-ray emission has been detected so far and where the density upper limit is very constraining
  \citep[< 0.02 cm$^{-3}$, ][]{cassam04}. Because of the density difference, we could have expected a higher fraction of hadronic emission in RCW 86 and thus a steeper spectral index at HE.
  We emphasize that this possible hadronic contribution on small scales is at the sensitivity and angular resolution limit of current generation instruments. 
We particularly anticipate the next generation of Cherenkov telescope CTA to carry out spatially resolved spectroscopy on a large energy range to test those 
predictions.

\section{Summary and conclusion}

Using 6 years of P7REP data of the \fermi telescope, we have studied the HE counterparts of the SNRs \src and SN 1006.
Although both objects are not detected at those energies, we report new  upper limits that can rule out, at a confidence level $> 5 \sigma$, 
a standard hadronic emission scenario ($s_{\rm p}=2.0$) as the main mechanism for HE and VHE \g-ray emission.
Given that there is no hint of detection in 6 years of data, we do not expect a detection with \fermi in the years to come.

With this study, we now have a complete view of HE and VHE emission of the five TeV shell SNRs. 
All objects show hard spectral indices at HE ($1.4 < \Gamma < 1.8$) that can simply be explained in a standard leptonic dominated scenario.
While the SNRs are from different type of SNe (core collapse and Type Ia), have ages ranging from 1 to 6 kyrs and are evolving in different ambient media, they all 
show a surprisingly similar HE and VHE luminosity.

We emphasize that the fact that the HE and VHE emissions are likely to be dominated by leptonic emission does not rule out efficient hadrons acceleration in those 
TeV shell SNRs. However due to the low-density ambient medium  on average, the hadrons do not encounter sufficiently high target densities to 
produce a level of hadronic flux that can compete with the leptonic emission from the whole SNR. We note that in certain localized regions of those SNRs,
 the shock probably encounters enhanced densities and that locally the hadronic contribution might become important.
 The next generation of Cherenkov telescopes such as CTA will provide the necessary angular resolution to carry out detailed spatially resolved spectroscopy, 
 which might unveil different emission mechanisms depending on the regions.
 We anticipate that this will bring a shift in paradigm from "Is the \g-ray emission leptonic or hadronic dominated ?"
  to "In which region of the SNR is the emission leptonic/hadronic dominated ?".
 This evolution might be comparable to the question about the nature of the X-ray emission of SN 1006 in the 80's \citep[see ][and reference therein]{becker80}
  which debated whether the emission was Crab-like (power-law) or Tycho-like (thermal).  The ASCA satellite with an improved energy range, sensitivity, and angular resolution revealed that both emission mechanisms co-exist but with a different spatial distribution \citep{koyama95}.

\begin{acknowledgements}
The \textit{Fermi}-LAT Collaboration acknowledges generous on- going support from a number of agencies and institutes that have supported both the development and the operation of the LAT as well as scientific data analysis. These include the National Aeronautics and Space Administration and the Department of Energy in the United States, the Commissariat \`a l'Energie Atomique and the Centre National de la Recherche Scientifique/Institut National de Physique Nucl\'eaire et de Physique des Particules in France, the Agenzia Spaziale Italiana, the Istituto Nazionale di Fisica Nucleare, and the Istituto Nazionale di Astrofisica in Italy, the Ministry of Education, Culture, Sports, Science and Technology (MEXT), High Energy Accelerator Research Organization (KEK) and Japan Aerospace Exploration Agency (JAXA) in Japan, and the K. A. Wallenberg Foundation and the Swedish National Space Board in Sweden. Additional support for science analysis during the operations phase from the following agencies is also gratefully acknowledged: the Instituto Nazionale di Astrofisica in Italy and the Centre National d'Etudes Spatiales in France. \end{acknowledgements}

\bibliographystyle{aa}

\end{document}